\documentclass[namedreferences]{solarphysics}
\usepackage[hyperref,optionalrh,showbiblabels]{spr-sola-addons} 
\usepackage{graphicx}        
\usepackage{color}           
\usepackage{breakurl}        

\chardef\us=`\_

\begin{document}
\begin{article}
\begin{opening}
\title{Analysis of EUV Solar Spectrum with \emph{gire}}
\author[addressref={aff1,aff2},corref,email={antonysoosaleon@yahoo.com}]{\inits{S. Antony}\fnm{Antony Soosaleon}}
\author[addressref=aff3,email={sunithaambrose@gmail.com}]{\inits{A. Sunitha}\fnm{Sunitha Ambrose}}
\address[id=aff1]{The Astronomical Observatory, Department of Physics, University of Kerala,  India - 695033 }
\address[id=aff2]{School of Pure \& Applied Physics, Mahatma Gandhi University, Kottayam, Kerala, India-686560}
\address[id=aff3]{Center for Fundamental Research and Computational Sciences,
Thiruvanathapuram, Kerala, India - 695043. }

\runningauthor{Antony et al.}
\runningtitle{Example paper}
\begin{abstract}
Gravity Induced Resonant Emission (\emph{gire}) is a new phenomenon recently reported \citep{Antony2017}, we have analyzed the solar emissions on the basis of \emph{gire}.  We have computed the EUV solar spectrum \citep{Aschwanden2005, Harrison1995} using \emph{gire} wave length as a part of studying coronal heating problem.  We find all the solar observations converge in \emph{gire} and the \emph{gire} wave length is sufficient to study all the solar emissions.
 
\end{abstract}
\end{opening}
\section{Introduction}
    \label{S-Introduction} 
The presence of highly ionized heavy elements in solar corona was noticed by \citep{Edlen1943, Edlen1945, Grotrian1939}, which led to the problem of coronal heating, one of the most haunting problem in solar physics for the last seven decades. Corona is the source of solar wind which directly affects the earth climate and space equipments also the quest for energy from fusion, compelled several curious minds in to this strange phenomenon. Untiring search of reasons by various missions for the last 7 decades made tremendous progress in this area. Numerous journal publications and books regarding the developments over this problem could be found \citep{Aschwanden2005, Kenneth2009} also in the website of NASA data system. One of the major data in the coronal heating problem is about the details of heavier elements present in the corona, detected by CDS, EIT and TRACE (CDS, Coronal Diagnostic Spectrometer; EIT, Extreme-ultraviolet Imaging Telescope; TRACE, Transition Region And Coronal Explorer (spacecraft)), compiled as EUV Solar Spectrum. As a part of studying coronal heating problem, on the basis of gravity induced resonant emission (\emph{gire}), we are analyzing the EUV spectrum and  this analysis is done in the following manner
\begin{itemize}
\item First we list the important characteristics of \emph{gire} wave length
\item Establishing these characteristics over the observational data about the solar emission
\item Generating EUV solar spectrum with \emph{gire} wave length
\item Analyzing the variables over the the generated EUV spectrum by \emph{gire}
\end{itemize}

\section{Characteristics of \emph{gire}}
 The effect of gravity is more on heavier ions relative to electrons induces an electric field; this electric field is very effective at high vacuum condition. Due to this electric field lower hybrid (\emph{lh}) oscillations are induced by the propagation of \emph{em} waves in the perpendicular direction. This \emph{lh} oscillation leads to the resonant absorption and re-emission of \emph{em} waves is very effective in ultra vacuum condition. This is named as Gravity Induced Resonant Emission (\emph{gire}). Solar corona is extremely vacuum; hence \emph{gire}may emerge as a most probable solution to Coronal Heating problem. The \emph{gire} frequency is given as $$ \omega_r^2 = \frac{c^2 \omega_{l}^2}{g} \frac{n_0^{'}}{n_0} $$ 
 Where $\omega_l = \frac{|eB|}{\sqrt{mM}}$ is the \emph{lh} frequency; e, charge; B, magnetic field; m, electron mass; M proton mass;  $ {n^{'}_{\circ}}$, is the number density of ions; $ {n_{\circ}}$ is the number density of electrons. The \emph{gire} frequency is deduced from the resonant absorption of \emph{em} waves in the \emph{lh} oscillation; to list the characteristics of \emph{gire}, we need to know the physics of cyclotron resonant damping in the \emph{lh} oscillations and also we should analyze this over solar coronal conditions.
 
The \emph{lh} oscillation is a coupled oscillation between the oppositely charge particles bound by Coulomb force when they are in cyclotron motion due to Lorentz force. The cyclotron motion of ions and electrons are in the opposite directions hence they mutually cross over by two times in every rotation in the \emph{lh} oscillation (see figure 1) if the frequencies are same. Here the particles are connected by the Coulomb force (attractive); when they move towards, they get accelerated and this acceleration occurs by the neutralization of their own fields (+ve \& -ve). Here the field energy of the particles is converted into kinetic energy and thus the energy is conserved. 

Now with the increased kinetic energy they cross over (unless the electron falls into the ion or they collide), they feel a sudden deceleration due to the deficit of fields since they are moving against the attractive force (see the figure 1). At this point, the particles are in the desperate need of field energy to compensate the lost field. At this situation, they absorb the necessary field energy from incoming \emph{em} waves to overcome the attractive force and move forward to meet at the other point. \textbf{Thus, in the first half of the orbital path, they convert their field energy in to kinetic energy by the neutralization of their own field; and in the other half, they absorb the field energy from the \emph{em} waves which they lost by field neutralization. Thus for a complete orbital motion, the particles convert the \emph{em} waves in to kinetic energy. This is one of the most beautiful mechanism of auto conversion of field energy in to kinetic energy in \emph{lh} oscillation.}

\subsection{Time Factor of Cyclotron Damping process in \emph{lh} oscillation}
A very important point in the \emph{lh} oscillation is the acceleration and deceleration times. As we have discussed earlier, the particles in the \emph{lh} oscillation; for half of the orbital length, the particles are accelerated towards by the Coulomb force and in the other half they are moving against the attractive force. We know, when the particles in the attractive force, the acceleration time ($t_a$) would be much smaller than the deceleration time ($t_d$, since when they move towards the velocities are added up). Here the electrons move in relativistic speeds, therefore; $t_d >> t_a$; this means, most of the time the particles are in the deceleration mode i.e., in the resonant absorption condition.\textbf{Thus the particles in the \emph{lh} oscillation, continuously absorbing the field energy from the \emph{em} waves and converting it into kinetic energy; since the solar atmosphere is sufficiently rich with all type of radiation, the gravity induced \emph{lh} oscillation is sufficient to heat ions to any required level of temperature.} 

\subsection{Orbital elongation and Field deficit in \emph{lh} oscillation}
Here one more thing to be noted is that the rate of absorption of the field energy is different for electrons and ions. The resonant absorption of energy by the particles due to the field deficit is directly depend on the cyclotron frequency. The electrons rotate with high frequency while the ions at lower rate, since $\omega_e > \Omega_i$, therefore the rate absorption of \emph{em} energy by the electrons are higher than the ions. The energy of the electrons increases first; then its Larmour radius increases to conserve the angular momentum. When the orbital radii of the electrons increases; this gives more acceleration time to the coupled ions and they are get accelerated to higher energy levels and hence the orbits also elongated. These elongations of orbits are mutually in the opposite directions because the ions are controlled by electric field (gravity induced) and the electrons by magnetic field (figure 2). In this situation if there is a change in the magnetic field affects the density of electrons rather the density of ions. We know the coronal magnetic field is highly dynamical and such dynamical oscillations of magnetic field is greatly modulated in the density of electrons which leads to the deficit of field to ions. And this field deficit is the basic mechanism of ion cyclotron resoanat damping in the \emph{lh} oscillation; this is the most important part of this paper, we shall give little more explanation on the basis of coronal oscillations. 

 \begin{figure}    
   \centerline{\hspace*{0.015\textwidth}
               \includegraphics[width=0.4\textwidth,clip=]{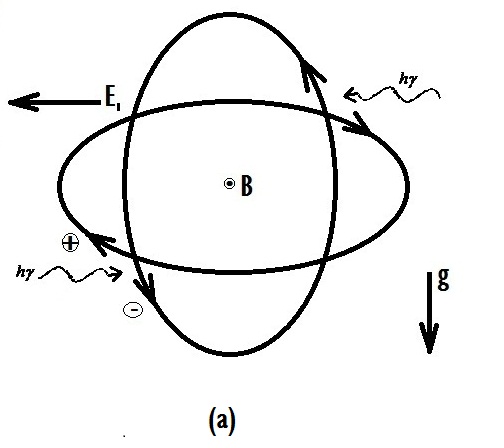}
               \hspace*{-0.03\textwidth}
               \includegraphics[width=0.615\textwidth,clip=]{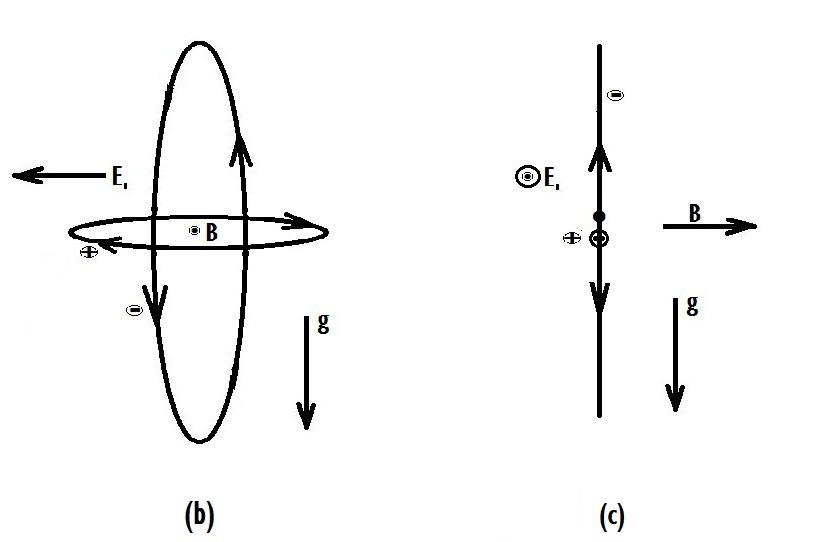}
              }
     \vspace{-0.35\textwidth}   
     \centerline{\Large \bf     
      \hspace{0.0 \textwidth}  \color{white}{(a)}
      \hspace{0.415\textwidth}  \color{white}{(b)}
         \hfill}
     \vspace{0.31\textwidth}    
              
\caption{(a) Photo Absorption at Field Deficit;  (b) Orbital elongation at high Energy, view parallel to B and (c) view $\bot$ to B (plain of rotation $\bot$ plain of paper)}
   \label{F-2panels}
   \end{figure}

\subsection {Coronal Oscillation and Electron Deficit} The oscillation of the coronal loop is a regular situation in the solar atmosphere. Here we are going to prove that this coronal oscillation is the secret of the mysterious coronal heating problem. The magnetic field in the solar atmosphere is produced by the convection currents (below the photosphere), which emerges from the foot loop of the photosphere and passes through the transition layer to corona and then bend back to another foot loop. Solar surface is highly turbulent due to pressure wave, which causes Kink and Sausage oscillations in the coronal loop. These oscillations passes through the loop and finally end up in the top of the loop purely as transverse oscillations (see figure 2a). These oscillations causes sudden reduction in the magnetic field spatially which is maximum at the top of the loop (where displacement is more and field strength is less). This reduction in the magnetic field intensity leads to the heavy loss of ionized particles and this loss is more to the electrons rather than to ions in the gravity induced \emph{lh} oscillations. This will be clear if we analyze the orbitals of the particles which undergo gravity induced \emph{lh} oscillations in the coronal loops. To see the actual situation in the loop, we shall super impose the orbitals of gravity induced \emph{lh} oscillations (fig.1b) over the coronal loop. The fig.1b, shows the orbitals of the particles where the direction of \textbf{B} is normal to the plane of the paper. But the direction of \textbf{B} in coronal loop is parallel to the plain of paper and therefore we rotate the orbitals $\frac{\pi}{2}$ to make B parallel to the plain of the paper and now the view of orbitals parallel to \textbf{B} is shown in figure 1c. Now we can super impose the fig.1c over the coronal loop oscillations (fig.2a) and the combined figure is shown as fig.2b (kindly note that the fig.2b is symbolic sketch and the orbital dimensions are much exaggerated to match with coronal loop, even otherwise the boundary of B is not definite in the particle dimension). 
  \begin{figure}    
   \centerline{\hspace*{0.015\textwidth}
               \includegraphics[width=0.515\textwidth,clip=]{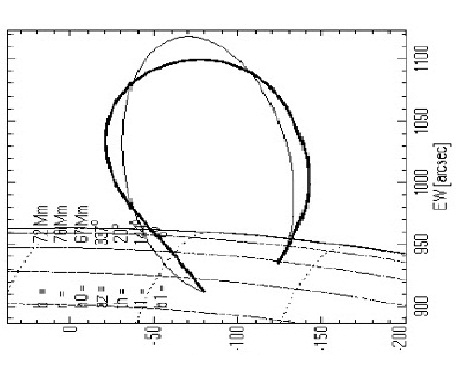}
               \hspace*{-0.03\textwidth}
               \includegraphics[width=0.515\textwidth,clip=]{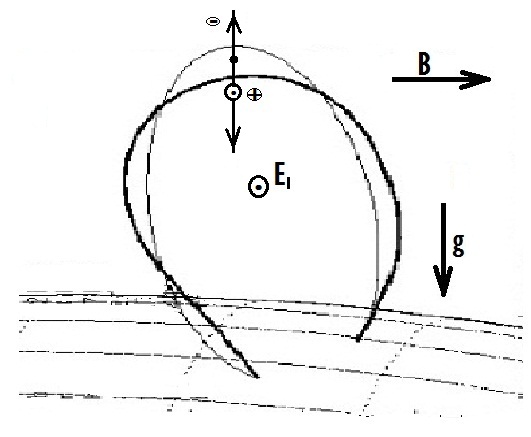}
              }
     \vspace{-0.35\textwidth}   
     \centerline{\Large \bf     
      \hspace{0.0 \textwidth}  \color{white}{(a)}
      \hspace{0.415\textwidth}  \color{white}{(b)}
         \hfill}
     \vspace{0.31\textwidth}    
              
\caption{(a) Model of Kink oscillation, Courtesy: Ashwandan M., (2005); (b) Process of electron density deficit in Kink Oscillation}
   \label{F-2panels}
   \end{figure}

Now it is easy to analyze from the fig.2b which depicts the real situation of particles motion in the \emph{lh} oscillation over the coronal loop. We find the coronal loop displacement is parallel to the elongation of the electrons and normal to the motion of ions. This shows the electrons in the gravity induced lower \emph{lh} oscillations have minimum critical thermal velocity could escape the Coulomb potential of ions. And this loss of electrons may depend on several factors like oscillation amplitude, time, \textbf{B}, Temperature and the density (more theory is needed). Thus the variation in the magnetic field does not affect the ions much but the electrons and this loss would be high for weak field situation (solar minimum, $B\approx 10-50$G). This shows that in each oscillation, millions of coupled electrons may be escaping from the coronal loop and create a heavy reduction in the density of electrons has been confirmed by several scientists \citet{Banerjee1998, Doschek1997, Spadaro1999, Pekeris1948, Murawski1998, Murawski1994, Murawski1993, Roberts1994} that the electron density decreases in a logarithmic scale. Thus the fluctuations in the coronal loop causes a great loss in the density of electrons, most of the magnetic field loops end in solar corona where the field deficit is very high and this is the reason that we notice the emission is along the magnetic loops. The acceleration and elongation of the orbits of the particles continued until the electrons escape from the Coulomb potential of the ions or they may under go a head on collision. If the electrons escape from the Coulomb potential of ions, this will increase the acceleration time for ions and the ions will be accelerated to high energy. Whether it is electron loss or the head on collision with ions; in both the situations, the emission is possible. 

\textbf{Now we understand that in the gravity induced \emph{lh} oscillations, the electrons and ions are confronting twice during every cycle of rotation when the frequencies are same or several times ( practically $\Omega_e >> \Omega_i$)  and therefore, \emph{gire} has the highest possibility of becoming the mother of all the atomic transitions (more work needed in the possibility of various transitions)}. \\

From the above discussions we can conclude that the Ion cyclotron damping in the lower hybrid oscillations is a continuous process and the absorption is maximum when the Larmour radius of ions and electrons are same ($r_{Li} = r_{Le}$. At this situations there are number of possible transitions could take place such Couloumb's Interaction which leads to momentary emission of radiation from the ion. The dispersion relation of a two component plasma, which has been derived by solving the momentum and continuity equations of MHD theory. The two components in the plasma are ions and electrons, therefor the ion can be any ionized heavier elements and the electron gives only the charge neutrality. For getting a generalized formula suitable to all ionized element, we replace $ \Omega_i  = \frac{Z}{\mu} \Omega_i $, where $\mu$, mass number; Z, degree of ionization. The general form of \emph{gire} wavelength is given as
 \begin{equation}
 {\lambda}_r = 2 \pi [(\frac{\mu}{Z}) (\frac{g}{\omega_l})(\frac{n_{\circ}}{{n_{\circ}}^{'}})] ^{\frac{1}{2}}
\end{equation}
Here we find the \emph{gire} wavelength depend on seven variables and each variable is very much sensitive to the coronal plasma and  we list  important characteristics of \emph{gire} wavelength below.

\begin{enumerate}
\item The \emph{gira} wave length is deduced by using Magneto Hydro Dynamic (MHD) theory
\item \emph{gira} is a non thermal process i.e., cyclotron resonance absorption by \emph{lh} oscillation
\item Resonant absorption in \emph{lh} oscillation depends on field deficit \cite{ref16}     
\item For ions the field deficit depend on the decrease in the electron density
\end{enumerate}
Now we will be discussing these listed properties with the established results of coronal emissions. 

\section {Observational Convergence over \emph{gire}} First we shall see, how MHD theory is suitable to solar physics problems. 
\subsection{MHD waves in UV Emissions} Solar atmosphere is highly dynamical and structured in density and fields, therefore the effect of  spatial and temporal oscillations of fields and density cannot be avoidable. Since the solar oscillations are macroscopic in nature, describing the macroscopic dynamical structures of magnetic field embodied with plasma, only MHD theory could help. This is the reason that the MHD oscillation has been observable in all range of wavelengths (for complete detail see Ashwanden 2005). This study is over the EUV spectrum, therefore we are mentioning the observation of EUV range only \citep{Antonucci1984, Aschwanden1999, Aschwanden2002, Chapman1972, Nakariakov1999, Nakariakov2001}, for many situations the EUV and soft X-rays are observed together.  The second characteristic of \emph{gire} is it is a non thermal process.

\subsection{MHD with Non Thermal Emission}The morphological changes in coronal magnetic field produces local heating reported by several scientists \citep{Tarbell1999, Tarbell2000, Ryutova2000, Ryutova2001}. Micro flares in the solar atmosphere causes MHD wave propagation which is modulated in to gyro resonance and non thermal micro wave emissions \citep{Gopalswamy1994, Gopalswamy1997, White1995, Gary1997, Nitta1997, Berghmans2001, Shimizu1992, Shimizu1994, Shimizu1995, Shimizu1997a, Shimizu1997b}. An analytical model on MHD turbulent showed that the turbulent velocities can be correlated with the excess broadening of lines observed by SUMER \citep{Heyvaerts1992, Inverarity1995a, Inverarity1995b, Milano1997}. Various numerical simulation pointed out that the nonlinear wave dissipation could be linked with the radio emission which are morphologically sensitive in nature\citep{Ofman1999, Drago1974, Lantos1975, Furst1975, Kundu1976, Chiuderi1977a, Chiuderi1977b, Chiuderi1999, Dulk1977, Papagiannis1982, Wang1987, Kundu1989, Bogod1997, Gopalswamy1999, Moran2001} and also in the study on turbulent heating, a high degree of spatial and temporal intermittent has been observed\citep{Einaudi1996a,Einaudi1996b, Dmitruk1997} and the frequency distribution of this simulation was found to be in correlation with the nano flare distribution\citep{Dmitruk1998}. Small scale phenomenons such as prominences and striations in the transition and coronal region exhibits various signatures of all range of non thermal emissions \citep{Parnell2002a, Parnell2002b, Shimizu2002a, Shimizu2002b, Berghmans2002}. Finally we look in to the figure showing the integrated Energy spectrum of flare electrons taken by RHESSI (Ramaty High Energy Solar Spectroscopic Imager) , NASA, on 20 February 2002, clearly indicate the emission spectrum is fully non thermal and the Hinode \citep{Zirker1993} observation confirmed the resonant absorption. 
 
\subsection{Evidence of lower hybrid Cylcotron Resonant Damping} \emph{lh} waves can simultaneously heat electrons and ions \citep{McBride1972}, such waves are excited when the ions are drifting perpendicular to the magnetic field, these distributions are found in space plasmas are arising due to diamagnetic drift associated with density gradients \citep{Krall1971}. In formulating a unified nonlinear theory for \emph{lh} waves, in the study of ion \& electron  damping process, it is found that \emph{lh} wave causes direct transfer of \emph{em} energy to both ion and electrons by damping process\citep{Tripathi1977, McClements1993}. There are several studies in which the \emph{lh} waves have been invoked to explain for the acceleration of electrons causing solar radio burst type I \citep{Spicer1981}, type II \citep{Lampe1977, Thejappa1987} type IV \citep{Vlahos1982}, also \citet{Benz1987} proposed that \emph{lh} waves could responsible for acceleration of flare electrons. 
  
\subsection{Evidence of Density Dependance} In the study of Sausage Mode oscillations, scientists found that the density variation in the electrons easily modulate in to gyro synchrotron emission \citep{Rosenberg1970, Ofman1997, Ofman2000a} and also it found that the electron density of the plasma reduces tremendously above the foot points where the heating is more by \textbf{TRACE}, \citep{Aschwanden2000,  Aschwanden2001a, Chae2002}. Study of polarized brightness over the coronal holes by Time Series Analysis revealed that the average density profile can be inverted with brightness \citep{Fisher1975, Guhathakurta1996} this is in correlation with the \textbf{TRACE} data of electron density reduction over foot point. In the study of soft X-ray emission by OSO-7 and Skylab showed very high density fluctuations \citep{Timothy1975, Hara1994, Hara1996, Hara1997, Watari1995, Foley1997, Aschwanden2001b} and gyro resonance emission has also been observed around a symmetric sun spots \citep{Alissandrakis1980, Krueger1985, Brosius1989, Lee1993a, Lee1993b, Vourlidas1997}. A study on the large flares confirmed the non thermal emission which reflected a density modulation in the EUV brightness variation \citep{Lin2001, DeForest1998}. Study over prominences by numerical simulation insisted a need of dip in the top of the horizontal magnetic field lines to explain the energy balance (see the figure 2) \citep{Kippenhahn1957, Pikelner1971, Orrall1961,Low1975a, Low1975b, Lerche1977, Heasley1976, Milne1979}, we suppose such dip could cause high electron deficit which lead to the high absorption of \emph{em} waves by \emph{gira}. From the above three subsections, we can arrive at the following conclusion;

\begin{enumerate}
\item The MHD waves are observable in all range of \emph{em} waves from Radio wave to Hard X ray and hence it would be the most suitable theory
\item Most of the emissions are non thermal confirms and gyro emission confirms the cyclotron resonant absorption is the true mechanism which accelerate all the ionized particles of corona.
\item Fluctuations in electron density inversely correlated with spectral emission (field deficit) confirms the field deficit is behind the spectral emission 
\item The \emph{lh} waves has been used to study of radio bursts to X-ray emissions shows the \emph{lh} waves are sufficient for the study of coronal heating
\end{enumerate}
All the observations and simulations converges in \emph{gire} which may be the ultimate mechanism behind the coronal emission. Now let us move to the next section of generating EUV solar spectrum on the basis of \emph{gira} wavelength.

\section{Computation EUV Spectrum} There are 6 variables in the expression of \emph{gire} wavelength (Eq.1), among these the most important variable is density ratio and the next critical variable is temperature, but it is not visible in the formula. 
\begin{table}
\caption{ Comparison of Calculated and Observed Temperatures}
\label{T-simple}
\begin{tabular}{cccccclc}     
\hline 
Line & Instr. & $T_{iO}$ &	$T_{iC}$	& $T_{eC}$ & $E_e (keV) $ & ${\gamma}$ \\
\hline 
He II$^*$ & CDS &	0.08 &	0.074	& 500 &	43.125 & 1.08 \\
O III	& CDS & 0.08 &	0.074 & 500 & 43.125 & 1.08  \\
O IV & CDS & 	0.20 & 	0.202	& 600	& 51.75 & 1.1	 \\
O V	& CDS & 0.25  & 	0.233	& 400	& 34.5 & 1.06  \\
Ne III & CDS &	0.08 &	0.072 &	600	& 51.75 & 1.1	\\
Ne V & CDS & 	0.30 & 	0.29 & 600 & 51.75 & 1.1	 \\
Ne VI & CDS & 0.40 & 	0.37 & 500	&	43.125 & 1.08 \\
Ne VII  & CDS & 0.50 &	0.42	& 400	& 34.5 & 1.06	 \\
Ne VIII & CDS &	0.63 &	0.57	& 400	& 34.5 & 1.06	 \\
N IV 	& CDS & 0.10 &	0.11  & 300 & 25.88 & 1.04	 \\
Mg VI & CDS &	0.40 &	0.375	& 600	& 51.75 & 1.1	 \\
Mg VII & CDS 	& 0.63 &	0.64	& 700	& 60.38 & 1.118	 \\
Mg VIII & CDS &	0.80 &	0.74	& 600	& 51.75 & 1.1	 \\
Mg IX & CDS  &	1.00 & 	0.96	& 600	& 51.75 & 1.1	 \\
Mg X 	& CDS  & 1.10  & 1.0	 & 500 &	43.125	& 1.08	 \\
Si VII & CDS  & 0.63 & 	0.64	& 800	& 69.00 & 1.135	 \\
Si VIII & CDS &	0.80 &	0.75 &	700	 & 60.38 & 1.118  \\
Si IX & CDS &	1.00 &	0.98	& 700	& 60.38 & 1.118	 \\
Si X 	& CDS & 1.00 &	1.04	& 600	& 51.75 & 1.1   \\
Si X 	& CDS & 1.30	& 1.23	& 700	& 60.38 & 1.118	 \\
Si XII & CDS 	& 1.80 &	1.84  &	700	& 60.38 & 1.118	 \\
Ca X 	 & CDS  & 0.63 	& 0.6	& 500	 &	43.125 & 1.084  \\
Al XI & CDS & 1.40 	& 1.28	& 700	& 60.38 & 1.118	 \\
Fe VIII & CDS &	0.40 &	0.373	& 700	& 60.38 & 1.118	 \\
Fe IX$^*$ & CDS & 1.00 &	0.99	& 1300 & 112.13	& 1.22 \\
Fe X	& CDS & 1.30 & 	1.25	& 1300	& 112.13 & 1.22	 \\
Fe XI & CDS & 1.30 &	1.27 &	1100 & 94.88 & 1.19	 \\
Fe XI & CDS & 1.30	& 1.27	& 1100	& 94.88 & 1.19	 \\
Fe XII$^*$ & CDS & 1.60 & 1.54	& 1100 & 94.88 &	1.19	 \\
Fe XIII & CDS &	1.60	& 1.64	& 1000	& 86.25 & 1.169	 \\
Fe XIV & CDS 	& 2.00 & 	1.92	& 1000 & 86.25 & 1.169	 \\
Fe XV$^*$ & CDS & 2.00 & 	1.98	& 900	& 77.63 & 1.15	 \\
Fe XVI 	& CDS & 2.50 & 	2.55	& 1000 & 86.25 &	1.169  \\
 \hline
\end{tabular}
\end{table}

\subsection{Temperature of Ionized Particles} Even though the \emph{gire} wavelength does not depend on temperature, but the information is contained in the relativistic factor ($\gamma$) of electron in $\omega_l$. Therefore the temperatures of the coupling electrons with the ions in the \emph{lh} oscillations to be found out for the calculation of $\gamma$. Here the electrons and ions are bound by Coulomb’s force undergoes cyclotron motion which are in opposite directions and the resonant emission occurs when particles are much closer such that the Larmour radius of electrons and the ions are same. Implying this condition we can deduce the following relation 
\begin{equation}
T_i = \frac{Z^2}{\mu }\frac{\gamma m}{M}T_e
\end{equation}  
Where $T_i$ and $T_e$ are the temperatures of ion and electrons respectively but here the problem is that the temperature of the electron again depend on $\gamma$. Therefore we design the program with floating variables to tune the relativistic term. This done by finding the coupling electron temperature (set as round off value to find the relativistic term) with the ion, which has the temperature approximately equal to the observed value of the ion temperature in the EUV spectrum. Due to this reason the theoretical value of the ion temperature has a finite difference with the observed value. Finding the temperature of the coupling electron corresponding to each ionized element, the relativistic term is fixed. We have computed the $T_i$, $T_e$ and $\gamma$ for all the coronal elements detected by CDS, *EIT and *TRACE \citep{Aschwanden2005} shown in the \textbf{Table.1}. This table contains four parameters: The first and second column shows the elements and their corresponding temperature ($T_{iO}$) in the observed EUV spectrum. The third column shows the values of possible ion temperature $T_{iC}$calculated theoretically which could under go \emph{gire}. The fourth column gives the value of the temperature of the electron $T_{eC}$ coupling with the ions in the LH oscillation. The last two column gives is the energy and relativistic factor of electron $\nu$ corresponding to the temperature $T_e$. This relativistic factor is used for the calculation of resonant wavelength shown in the table 2.\\

\subsection{Electron Density Reduction Factor} Our formula has got the ratio of the ion to electron density, they should be fixed on the basis of the observational data. As we see in the section 2.2, that the electron density decreases due to the coronal oscillations in a logerithemic scale.  The observations of \textbf{SoHO/SUMER} reveals, in the height range 20Mm to 180 Mm, the density of electron decreases approximately from ${1.1\times 10^{14} m^{-3}} $ to $1.6 \times{10^{13}m^{-3}} $\citep{Spadaro1990a, Spadaro1990b, Doyle1998, Erdelyi1998, Chae1998, Esser1999, Doschek2001}. Thus the change in electron density over a distance of of 160 Mm, is $ 15 \times 10^{14}m^{-3}$ and therefore the electron density reduction factor (EDRF) per meter is $\frac{15 \times 10^{14}}{16 \times 10^7 m^{-3}} \approx 1 \times 10^7$ (MKS units are used for computation). This shows the electron density ($n_{\circ}$) decreases logarithmically by a factor of 7. Now to arrive at the value of $\frac{n^{'}_{\circ}}{n_{\circ}}$ for the calculation and we are not certain about the value of ion density ($n^{'}_{\circ}$) (could be possible to find for a specific spectral line), we keep the value of $\frac{n^{'}_{\circ}}{n_{\circ}}$ as a floating variable in a logarithmic scale between $10^0$ to $10^7$, it could be even more these values, since the solar atmosphere is highly unpredictable.\\
\begin{table}
\caption{Comparison of Observed and Calculated Wavelengths}
\label{T-simple}
\begin{tabular}{cccccclc}     
\hline 
Line &  Instr. & $\lambda_O (A^{\circ})$ & $\lambda_C$ & B(G) & $(\frac{{n_{\circ}}^{'}}{n_{\circ}})$ & $\theta(^{\circ})$\\
\hline 
He II$^*$  & CDS & 303.78 &	303.61	& 12 & $10^6$  & 73 \\
O III & CDS	 &  702.98 &	704.12 & 23 &  $10^5$ & 71 \\
O III	& CDS  &  599.59 & 	599.07	& 26 &  $10^5 $ & 61 \\
Ne III  & CDS &	489.50 &	488.91	& 12 &  $10^6$ & 75 \\
N IV 	& CDS  & 765.14 &	765.12 &	16 &  $10^5$ & 74 \\
O IV 	& CDS  &  554.52	& 554.32	& 24 &  $10^5$ & 70 \\
O V	& CDS	 &  172.17 & 172.52	& 20 & $10^6$ & 61 \\
O V	& CDS	 &  629.73 &	628.51 &	17 &  $10^5$ & 57 \\
Ne V   & CDS &  482.10 & 	482.13	& 27 &  $10^5$ & 72 \\
Ne VI & CDS  & 399.83 & 399.95	& 29  &  $10^5$ & 74 \\
Ne VI & CDS  &  401.14 &	400.92 &	29 &  $10^5$ & 75 \\
Ne VI & CDS  &  562.83 	& 562.95	& 21 &  $10^5$ & 87 \\
Ne VI & CDS  &  558.59	& 558.14	& 21 & $10^5$ & 79 \\
Mg VI  & CDS &399.20 & 398.97 &	10 & $10^6$ & 68 \\
Mg VI  & CDS &	400.68 &	400.34 &	10 & $10^6$ & 69 \\
Fe VIII  & CDS &	168.18 & 	168.20	& 10 & $10^7$ & 67 \\
Fe VIII  & CDS &	186.60 	& 186.18	& 9 & $10^7$ & 66 \\
Ne VII  & CDS &  465.22 & 465.17	& 23 & $10^5$ & 90 \\
Ne VIII  & CDS &	 770.40	& 770.08	& 11 & $10^5$ & 47 \\
Ne VIII  & CDS &	780.30 &	779.15 &	12 & $10^5$  & 63\\
Mg VII  & CDS &  277.04	& 277.93 & 11 & $10^6$ & 40 \\
Mg VII & CDS &  278.40 	& 278.12	& 12 & $10^6$  & 50\\
Si VII 	& CDS &  272.60	& 272.80	& 15 & $10^6$  & 79 \\
Si VII  & CDS & 275.37 	& 275.34	& 15 & $10^6$ & 90 \\
Ca X  & CDS & 557.76 &	557.41 &	22 & $10^5$ & 76 \\
Mg VIII & CDS  & 313.73	& 313.58	& 11 & $10^6$ & 76 \\
Mg VIII & CDS &	317.01 &	317.16	& 11 & $10^6$ & 83 \\
Si VIII  & CDS &	316.22 & 	316.51 & 11 & $10^6$ & 56 \\
Si VIII & CDS  &	319.83 &	318.65	& 12 & $10^6$ & 90 \\
Fe IX$^*$  & CDS &  171.07 & 	171.1	& 10 & $10^7$  & 86 \\
Fe IX  & CDS &  217.10	& 217.34	& 24 & $10^6$  & 68 \\
Mg IX & CDS &  368.06 &	368.88 &	9 & $10^6$  & 64 \\
Mg IX  & CDS & 705.80 & 704.38	& 14 & $10^5$ & 65 \\
Si IX  & CDS &  296.12	& 295.54	& 11 & $10^6$ & 57 \\
Si IX & CDS  &	 345.12 &	345.19	& 10 & $10^6$ & 71 \\
Si X 	& CDS & 261.06	& 261.14 &	12 & $10^6$ & 63 \\
Si X 	& CDS & 	271.99	& 271.90	& 12 & $10^6$ & 75 \\
Mg X  & CDS &  624.94 	& 625.59	& 14 & $10^5$ & 55 \\
Si IX  & CDS & 341.94 &	341.87 &	10 & $10^6$ & 66 \\
Si X 	& CDS  &  347.40	& 347.78	& 9 & $10^6$  & 61 \\
Si X 	& CDS  &  356.01 & 	356.79	& 9 & $10^6$ & 67 \\
Fe X	& CDS  &  174.53 &	174.5	& 9 & $10^7$ & 71 \\
Fe X	 & CDS & 177.24 & 177.14	& 9 & $10^7$ & 77 \\
Fe XI  & CDS & 358.62 & 358.98	& 13 & $10^6$ & 72 \\
Fe XI	& CDS  & 180.40	& 180.68	& 25 & $10^6$ & 63 \\
Fe XI	& CDS  & 188.22 & 188.21	& 24 & $10^6$ & 63 \\
Fe XI  & CDS &  192.81 & 	192.65	& 24 & $10^6$ & 69  \\
Al XI  & CDS & 568.12	& 568.96	& 17 & $10^5$ & 75 \\
Al XI  & CDS &  550.03 & 	549.68	& 16 & $10^5$ & 53\\
Fe XII$^*$  & CDS & 195.12 &	195.42 &	21 & $10^6$ & 54 \\
Fe XII  & CDS &	193.51 &	193.46 & 23 & $10^6$ & 72 \\
Fe XII & CDS  & 364.47 &	364.79 &	12 & $10^6$ & 67 \\
Fe XII  & CDS & 346.85 &	346.44 &	13 & $10^6$ & 77 \\
Fe XIII  & CDS &	203.79 & 	203.91	& 21 & $10^6$ & 78 \\
Fe XIII  & CDS &	213.77 & 	213.69 &	20 & $10^6$ & 77 \\
Fe XIII  & CDS &	320.80 &	320.69	& 13 & $10^6$ & 68 \\
Fe XIII  & CDS & 359.64 &	359.45	& 12 & $10^6$ & 83 \\
Fe XIII & CDS &  348.18 & 348.61	& 12 & $10^6$ & 69\\
Si XII  & CDS & 520.66 & 520.4	& 18  & $10^5$ & 73 \\
Fe XIV & CDS  & 211.32 & 211.54 &	19 & $10^6$ & 69 \\
Fe XIV  & CDS & 220.08 & 220.09	& 17 & $10^6$ & 54 \\
Fe XIV  & CDS & 334.17	& 334.94 &	12 & $10^6$ & 69 \\
Fe XIV  & CDS & 353.83 & 353.66	& 11 & $10^6$ & 61 \\
Fe XV$^*$  & CDS & 284.16 & 284.60	& 13 & $10^6$ & 60 \\
Fe XV  & CDS & 327.02 & 327.04	& 12 & $10^6$ & 77 \\
Fe XVI & CDS  & 200.80 & 200.77 &	11 & $10^6$ & 76 \\
Fe XVI& CDS & 335.40 & 335.15	& 11 & $10^6$ & 65 \\
Fe XVI & CDS & 360.76 	& 360.38 & 10 & $10^6$ & 60 \\
 \hline
\end{tabular}
\end{table}
\subsection{Modification of formula suitable to Corona}
 Before going to the computation of wavelength, there are two important corrections to be made in the formula of \emph{gire} wavelength on the basis of coronal emissions. First one is the theoretical assumption $g \bot B$, this gives the drift velocity $v_0 = \frac{g}{\Omega_i}$. This condition is satisfied only at the top of the coronal loop, but practically the emissions are observable through out the loop. This shows that the gravitational drift is effective to all parts of the loop, since the effect of gravity depend on the geometry of the loop and this can be  rectified by introducing an angle $\theta$ between g \& B. Now $g\times B = gB Sin \theta$, by replacing g with $g Sin\theta$, we get the correct formula suitable to all parts of the coronal loop. The second one is that we have assumed $T_e = T_i$, but from table 1 we find $T_e >> T_i$. By imposing these two conditions, the formula will be modified as 
 \begin{equation}
 \lambda_r = [(\frac{\sqrt{2}\pi^2  mMg}{e^2}) (\frac{\gamma \mu}{Z}) (\frac{Sin\theta}{B^2}\frac{n_{\circ}}{{n_{\circ}}^{'}})]^{\frac{1}{2}}
  \end{equation}
Now the first bracket is the constant, the second bracket is fixed for particular ionized element and the third bracket has 3 variables which are  B, Sin$\theta$ and $\frac{n_{\circ}}{{n_{\circ}}^{'}}$. As we have already discussed that these 3 variables are highly tunable to coronal plasma and could be studied according to the condition for solar minimum or maximum. At high magnetic field intensity (solar maximum) the spectral resolution will be good and tuning will be very easy to the maximum level of accuracy but at solar minimum, the value of B is very less and hence the spectral resolution will be affected. Therefore we  choose solar minimum condition($B =10-50G$, rest all constants have usual value in SI units and $ g = 274ms^{-1}$), but we have an another variable $\theta$ and  therefore it could be possible to tune it to more correct value. For each element, the mass number ($\mu$) and degree of ionization(Z) are constant; $\gamma$, relativistic factor of coupling electron corresponding to the element is taken from the table 1. The charge state of elements are taken as the value less than one of the Roman numerals of ionized elements; for example: $ Fe XV \rightarrow Z = 14$.The Computed wavelength ($\lambda_C$) by \emph{gire} formula in comparison with the observed wavelength ($\lambda_O$) is shown in the table 2.

\section{Analysis of Parameters over wavelength} The resonant wavelength is depend on 8 parameters, the influence of each parameter is very elaborate but we shall most relevant parameters to the corona. 
  
\subsection{Effect of $\theta$} $\theta$ is the angle between g \& B. We find $\lambda_r \propto \sqrt{g Sin\theta}$; when $\theta$ is low at the bottom side of the loop where B is high. It is simple that when B is high, $r_L$ is small. Resonant wavelength, $ \lambda \propto r_L $; therefore the ions resonate with shorter wave length. This shows that at smaller angle or at the bottom side of the loop, shorter wavelengths of radiation will be absorbed and the rate of resonant absorption would be high and this is the reason that the HARD X-rays are observed more at the bottom of the coronal loop. 
   
\subsection{Magnetic field (B)} $\lambda_r \propto \frac{1}{B}$, for high field region, $\Omega$ is high and $r_L$ is very low, the wavelength will shorter. The reason is that the high magnetic intensity could hold relatively high energetic electrons ($r_L \propto \frac{v_ti}{B}$) and hence the coupling is possible highly relativistic electrons. This gives a reason for the presence of highly ionized elements are observed at high field region. As we have seen from the table, the heavier ions need high relativistic electrons to couple in the LH oscillation, which is possible in the high magnetic field regions. This situations are more realistic during solar maximum and the higher elements could be observable at the top of the loop during solar maximum; in solar minimum the higher elements would be more probable at the rear side of the loop where the magnetic field intensity is high. 

\subsection{Density Ratio $ \frac{{n_{\circ}}^{'}}{n_{\circ}}$} $\lambda_r \propto \frac{{n_{\circ}}^{'}}{n_{\circ}}$. Here the electron density, ${n_{\circ}}$ increases the wavelength. This is nothing other than what we have explained in the section EDRF, in the \emph{lh} oscillation, the absorption of energy solely depend on the field deficit. When ${n_{\circ}}$ increases the field deficit is less, absorption rate is low and hence resonance shifts to longer wavelength. For higher ion density, the relative electron density decreases and wavelength will be shorter. More work could be done by the observational solar physicists to evaluate this factor very precisely on the basis of the elemental abundance.\\

The analysis of other parameters like Z, degree of Ionization; $\gamma$, relativistic factor; $\mu$, mass number; would lead to the elemental abundances and FIP factor and that we reserve for another discussion.  \\

\section{Conclusion}
\begin{itemize}
	\item In corona; magnetic field pressure much greater than the particle pressure and therefore the magnetic field is controlling the motion of the particles. 
	\item If any particles becomes neutral by collision, ans such particle will be transparent to the magnetic field descends down; and this is the reason that the solar Corona is fully ionized.
	\item Solar Corona is ultra vacuum and the fields are long range; since the corona is fully ionized, all the particles are in the cyclotron motion must be attracted by Coulomb force which leads to the coupling of ionized particles in the lower hybrid oscillations. 
	\item Magnetic field in the corona is highly oscillatory since it connected with convection layer and this oscillations of magnetic field results in the loss of ionized particles lead to the field deficit. 
	\item As per the theory of gravity induced  \emph{lh} oscillations, the particles loss is very much high for electrons compared with ions and hence Ion Cyclotron Resonant Damping is efficient and continuous in the solar corona.
\end{itemize} 
 \textbf{We have approached the problem with the prevailing situation of the solar atmosphere and hence the gravity induced lower hybrid Ion cyclotron resonant heating is an unavoidable solution to the coronal heating problem}.  \\
\textbf{Acknowledgment}
This first author acknowledges the funding of UGC under Research Award Scheme (2016-2018). 

\end{article} 

\begin{thebibliography}{200}

\bibitem[\protect\citeauthoryear{Alissandrakis, Kundu \& Lantos}{1980}]{Alissandrakis1980}
Alissandrakis, C.~E., Kundu, M.~R., Lantos, P., 1980, A\&A 82, 30.

\bibitem[\protect\citeauthoryear{Antonucci, Gabriel \& Patchett}{1984}]{Antonucci1984}
Antonucci, E., Gabriel, A.~H., Patchett, B.~E., 1984, SP, 93, 85.

\bibitem[\protect\citeauthoryear{Antony}{2017}]{Antony2017}
Antony, S., Sunitha, A., 2017, arXive:1704.07225v1[physics.plasm-ph]. 

\bibitem[\protect\citeauthoryear{Aschwanden et al.}{1999}]{Aschwanden1999}
Aschwanden, M.~J., Newmark, J.~S., Delaboudiniere, J.~P., Neupert, W.~M., Klimchuk, J.~A., Gary, G.~A., Portier-Fornazzi, F.,  Zucker, A., 1999, ApJ, 515, 842.

\bibitem[\protect\citeauthoryear{Aschwanden, Nightingale \& Alexander}{2000}]{Aschwanden2000}
Aschwanden, M.~J., Nightingale, R.~W., Alexander, D., 2000, ApJ, 541, 1059.

\bibitem[\protect\citeauthoryear{Aschwanden \& Acton}{2001a}]{Aschwanden2001a}
Aschwanden, M.~J., Acton, L.~W., 2001a, ApJ, 550, 475.

\bibitem[\protect\citeauthoryear{Aschwanden}{2001b}]{Aschwanden2001b}
Aschwanden, M.~J., 2001b, ApJ, 559, L171.

\bibitem[\protect\citeauthoryear{Aschwanden}{2002}]{Aschwanden2002}
Aschwanden, M.~J. 2002, ApJ, 580, L79.

\bibitem[\protect\citeauthoryear{Aschwanden}{2005}]{Aschwanden2005}
 Aschwanden, M.~J., 2005, Physics of Solar Corona, Springer, Praxis Publishing, UK

\bibitem[\protect\citeauthoryear{Banerjee, Teriaca \& Doyle}{1998}]{Banerjee1998}
Banerjee, D., Teriaca, L., Doyle, J.~G., 1998, A\& A 339, 208.

\bibitem[\protect\citeauthoryear{Benz \& Wentzel}{1981}]{Benz1981}
 Benz, A.~O., Wentzel, D.~ G., 1981, A\&A, 94, 100-108.

\bibitem[\protect\citeauthoryear{Benz \& Smith}{1987}]{Benz1987}
Benz, A.~O., Smith, D.~F., 1987, SP, 107, 299.

\bibitem[\protect\citeauthoryear{Berghmans}{2002}]{Berghmans2002}
Berghmans, D., 2002, ESA, SP-506, 501.

\bibitem[\protect\citeauthoryear{Berghmans, McKenzie \& Clette}{2001}]{Berghmans2001}
Berghmans, D., McKenzie, D., Clette, F., 2001, A\&A, 369, 291.

\bibitem[\protect\citeauthoryear{Bogod \& Grebinskij}{1997}]{Bogod1997}
Bogod V.M. \& Grebinskij A.S., 1997, SP 176, 67.

\bibitem[\protect\citeauthoryear{Boyd and Sanderson}{2003}]{Boyd2003}
 Boyd, T.J.~M., Sanderson, J.~J.,  2005, The Physics of Plasmas, Cambridge University Press

\bibitem[\protect\citeauthoryear{Brosius \& Holman}{1989}]{Brosius1989}
Brosius J.W. \& Holman G.D., 1989, ApJ 342, 1172.

\bibitem[\protect\citeauthoryear{Chae, Sch¨uhle \& Lemaire}{1998}]{Chae1998}
Chae, J.~C., Schuhle, U., Lemaire, P., 1998, ApJ, 505, 957.

\bibitem[\protect\citeauthoryear{Chae, Poland \& Aschwanden}{2002}]{Chae2002}
Chae, J.~C., Poland, A.~I., Aschwanden, M.~J., 2002, ApJ, 581, 726.

\bibitem[\protect\citeauthoryear{Chapman, Jordan \& Neupert}{1972}]{Chapman1972}
Chapman, R.~D., Jordan, S.~D., Neupert, W.~M., 1972, ApJ, 174, L97.

\bibitem[\protect\citeauthoryear{Chiuderi, Avignon \& Thomas}{1977a}]{Chiuderi1977a}
Chiuderi, D.~F., Avignon, Y., Thomas, R.~J., 1977a, SP, 51, 143.

\bibitem[\protect\citeauthoryear{Chiuderi\& Poletto}{1977b}]{Chiuderi1977b}
Chiuderi, D.~F., Poletto, G., 1977b, A\&A, 60, 227.

\bibitem[\protect\citeauthoryear{Chiuderi, Landi \& Fludra}{1999}]{Chiuderi1999}
Chiuderi, D.~F., Landi, E., Fludra, A., 1999, A\&A, 348, 261.

\bibitem[\protect\citeauthoryear{DeForest \& Gurman}{1998}]{DeForest1998}
DeForest, C.~E., Gurman, J.~B., 1998, ApJ, 501, L217.

\bibitem[\protect\citeauthoryear{Dmitruk \& Gomez}{1997}]{Dmitruk1997}
Dmitruk, P., Gomez, D.~O., 1997, ApJ, 484, L83.

\bibitem[\protect\citeauthoryear{Dmitruk, Gomez \& DeLuca}{1998}]{Dmitruk1998}
Dmitruk, P., Gomez, D.~O., DeLuca, E.~E., 1998, ApJ, 505, 974.

\bibitem[\protect\citeauthoryear{Doschek, Warren \& Laming}{1997}]{Doschek1997}
Doschek, G.~A., Warren, H.~P., Laming, J.~M., 1997, ApJ, 482, L109.

\bibitem[\protect\citeauthoryear{Doschek, Warren \& Laming}{2001}]{Doschek2001}
Doschek, G.~A., Feldman, U., Laming, J.~M., 2001, ApJ, 546, 559.

\bibitem[\protect\citeauthoryear{Doyle, Banerjee \& Perez}{1998}]{Doyle1998}
Doyle, J.~G., Banerjee, D., Perez, M.~E., 1998, Solar Phys., 181, 91.

\bibitem[\protect\citeauthoryear{Drago}{1974}]{Drago1974}
Drago, F., 1974, Proc-1974-Righini, 120.

\bibitem[\protect\citeauthoryear{Dulk, Sheridan \& Smerd}{1977}]{Dulk1977}
Dulk, G.~A., Sheridan, K.~V.,  Smerd, S.~F., 1977, SP, 52, 349.

\bibitem[\protect\citeauthoryear{Edlen}{1943}]{Edlen1943}
Edlen, B., 1943, Z. Astrophysik, 22, 30.

\bibitem[\protect\citeauthoryear{Edlen}{1945}]{Edlen1945}
Edlen, B., 1945, MNRAS, 105, 323.

\bibitem[\protect\citeauthoryear{Einaudi, Velli \& Politano}{1996a}]{Einaudi1996a}
Einaudi, G., Velli, M., Politano, H., 1996a, ApJ, 457, L113.

\bibitem[\protect\citeauthoryear{Einaudi, Califano \& Chiuderi}{1996b}]{Einaudi1996b}
Einaudi, G., Califano, F., Chiuderi, C., 1996b, ApJ, 472, 853.

\bibitem[\protect\citeauthoryear{Esser, Fineschi \& Dobrzycka}{1999}]{Esser1999}
Esser, R., Fineschi, S., Dobrzycka, D., 1999, ApJ, 510, L63.

\bibitem[\protect\citeauthoryear{Erdelyi, Doyle \& Perez}{1998}]{Erdelyi1998}
Erdelyi, R., Doyle, J.~G., Perez, M.~E., 1998, A\&A, 337, 287.

\bibitem[\protect\citeauthoryear{Fisher \& Musman}{1975}]{Fisher1975}
Fisher, R.~R., Musman, S., 1975, ApJ, 194, 801.

\bibitem[\protect\citeauthoryear{Foley, Culhane \& Acton}{1997}]{Foley1997}
Foley, C.~R., Culhane, J.~L., Acton, L.~W., 1997, ApJ, 491, 933.

\bibitem[\protect\citeauthoryear{Furst \& Hirth}{1975}]{Furst1975}
Furst, E., Hirth, W., 1975, SP, 42, 157.

\bibitem[\protect\citeauthoryear{Gary, Hartl \& Shimizu}{1997}]{Gary1997}
Gary, D.~E., Hartl, M.~D., Shimizu, T., 1997, ApJ, 477, 958.

\bibitem[\protect\citeauthoryear{Gopalswamy, Payne \&  Schmahl}{1994}]{Gopalswamy1994}
Gopalswamy, N., Payne, T.~E.~W., Schmahl, E.~J., 1994, ApJ, 437, 522.

\bibitem[\protect\citeauthoryear{Gopalswamy, Zhang \&  Kundu}{1997}]{Gopalswamy1997} 
Gopalswamy, N., Zhang, J.,  Kundu, M.~R., 1997, ApJ, 491, L115.

\bibitem[\protect\citeauthoryear{Gopalswamym, Shibasaki\& Thompson}{1999}]{Gopalswamy1999}
Gopalswamy, N., Shibasaki, K., Thompson, B.~J., 1999, JGR, 104/A5, 9767.

\bibitem[\protect\citeauthoryear{Grotrian}{1939}]{Grotrian1939}
Grotrian, W., 1939, Naturwissenschaften, 27, 214.

\bibitem[\protect\citeauthoryear{Guhathakurta, Fisher \& Strong}{1996}]{Guhathakurta1996}
Guhathakurta, M., Fisher, R., Strong, K., 1996, ApJL, 471, L69.

\bibitem[\protect\citeauthoryear{Hara, Tsuneta\& Acton}{1994}]{Hara1994}
Hara, H., Tsuneta, S., Acton, L.~W., 1994, PASJ, 46, 493.

\bibitem[\protect\citeauthoryear{Hara, Tsuneta \& Acton} {1996}]{Hara1996}
Hara, H., Tsuneta, S., Acton, L.~W., 1996, Adv. Space Res., 17/4-5, 231.

\bibitem[\protect\citeauthoryear{Hara}{1997}]{Hara1997}
Hara, H., 1997, PASJ, 49, 413.

\bibitem[\protect\citeauthoryear{Harrison et. al}{1995}]{Harrison1995}
Harrison, R.~A., Bryans, P., Simnett, G.~M., Lyons, M., 1995, SP, 162, 233

\bibitem[\protect\citeauthoryear{Heasley \& Mihalas}{1976}]{Heasley1976}
Heasley, J.~N., Mihalas, D., 1976, ApJ, 205, 273.

\bibitem[\protect\citeauthoryear{Heyvaerts \& Priest}{1992}]{Heyvaerts1992}
Heyvaerts, J., Priest, E.~R., 1992, ApJ, 390, 297.

\bibitem[\protect\citeauthoryear{Inverarity, Priest\& Heyvarts}{1995a}]{Inverarity1995a}
Inverarity, G.~W., Priest, E.~R., Heyvarts, J., 1995a, A\&A, 293, 913.

\bibitem[\protect\citeauthoryear{Inverarity \& Priest}{1995b}]{Inverarity1995b}
Inverarity, G.~W., Priest, E.~R., 1995b, A\&A, 296, 395.

\bibitem[\protect\citeauthoryear{Kenneth}{2009}]{Kenneth2009}
Kenneth, R.~L., 2009, Springer Verlag Berlin Heidelberg.

\bibitem[\protect\citeauthoryear{Kippenhahn \& Schl¨uter}{1957}]{Kippenhahn1957}
Kippenhahn, R., Schl¨uter, A., 1957, Z.Astrophys. 43, 36.

\bibitem[\protect\citeauthoryear{Krall \& Liewer}{1971}]{Krall1971}
Krall, N.~A., Liewer, P.~C., 1971, Phys. Rev, A4, 2094

\bibitem[\protect\citeauthoryear{Krueger, Hildebrandt \& Fuerstenberg}{1985}]{Krueger1985}
Krueger, A., Hildebrandt, J., Fuerstenberg, F., 1985, A\&A, 143, 72

\bibitem[\protect\citeauthoryear{Kundu \& Liu}{1976}]{Kundu1976}
Kundu, M.~R., Liu, S.~Y., 1976, SP, 49, 267.

\bibitem[\protect\citeauthoryear{Kundu, Schmahl \& Gopalswamy}{1989}]{Kundu1989}
Kundu, M.~R., Schmahl, E.~J., Gopalswamy, N., 1989, Adv. Space Res., 9, 41.

\bibitem[\protect\citeauthoryear{Lampe\& Papadopoulos}{1977}]{Lampe1977}
Lampe, M., Papadopoulos, K., 1977, ApJ, 212, 886.

\bibitem[\protect\citeauthoryear{Lantos \& Avignon}{1975}]{Lantos1975}
Lantos, P., Avignon, Y., 1975, A\& A 41, 137

\bibitem[\protect\citeauthoryear{Lee, Hurford \& Gary}{1993a}]{Lee1993a}
Lee, J.~W., Hurford, G.~J., Gary, D.~E., 1993a, SP, 144, 45.

\bibitem[\protect\citeauthoryear{Lee, Hurford \& Gary}{1993b}]{Lee1993b}
Lee, J.~W., Hurford, G.~J., Gary, D.~E., 1993b, SP, 144, 349.

\bibitem[\protect\citeauthoryear{Lerche \& Low}{1977}]{Lerche1977}
Lerche, L., Low, B.~C., 1977, SP, 53, 385.

\bibitem[\protect\citeauthoryear{Lin, Feffer\& Schwartz}{2001}]{Lin2001}
Lin, R.~P., Feffer, P.~T., Schwartz, R.~A., 2001, ApJ, 557, L125.

\bibitem[\protect\citeauthoryear{Low}{1975a}]{Low1975a}
Low, B.~C., 1975a, ApJ, 197, 251.

\bibitem[\protect\citeauthoryear{Low}{1975b}]{Low1975b}
Low, B.~C., 1975b, ApJ, 198, 211.

\bibitem[\protect\citeauthoryear{McBride et. al}{1972}]{McBride1972}
McBride, J.~B., Otto, E., Boris, J.~P., Oran., J.~H., 1972, Phys. Fluids, 15, 2367

\bibitem[\protect\citeauthoryear{McClements et. al}{1993}]{McClements1993}
McClements, K.~G., Bingham, R., Su, J.~J., Dawson, J.~M., Spicer, D.~S., 1993, ApJ, 409, 465.

\bibitem[\protect\citeauthoryear{Milano, Gomez \& Martens}{1997}]{Milano1997}
Milano, L.~J., Gomez, D.~O., Martens, P.~C.~H., 1997, ApJ, 490, 442.

\bibitem[\protect\citeauthoryear{Milne, Priest \& Roberts}{1979}]{Milne1979}
Milne, A.~M., Priest, E.~R., Roberts, B., 1979, ApJ, 232, 304.

\bibitem[\protect\citeauthoryear{Moran, Gopalswamy\& Dammasch}{2001}]{Moran2001}
Moran, T.~G., Gopalswamy, N., Dammasch, I.~E., 2001, A\&A, 378, 1037.

\bibitem[\protect\citeauthoryear{Murawski, Aschwanden \& Smith}{1998}]{Murawski1998}
Murawski, K., Aschwanden, M.~J., Smith, J.~M., 1998, SP, 179, 313.

\bibitem[\protect\citeauthoryear{Murawski\& Roberts}{1993}]{Murawski1993}
Murawski, K., Roberts, B., 1993, SP, 144, 101.

\bibitem[\protect\citeauthoryear{Murawski \& Roberts}{1994}]{Murawski1994}
Murawski, K., Roberts, B., 1994, SP, 151, 305.

\bibitem[\protect\citeauthoryear{Nakariakov, Ofman \& DeLuca}{1999}]{Nakariakov1999}
Nakariakov, V.~M., Ofman, L., DeLuca, E., 1999, Science, 285, 862.

\bibitem[\protect\citeauthoryear{Nakariakov \& Ofman}{2001}]{Nakariakov2001}
Nakariakov, V.~M., Ofman, L. 2001, AA, 372, L53.

\bibitem[\protect\citeauthoryear{Nitta}{1997}]{Nitta1997}
Nitta, N., 1997, ApJ, 491, 402.

\bibitem[\protect\citeauthoryear{Ofman, Romoli \& Poletto}{1997}]{Ofman1997}
Ofman, L., Romoli, M., Poletto, G., 1997, ApJ, 491, L111.

\bibitem[\protect\citeauthoryear{Ofman, Romoli\& Poletto}{2000a}]{Ofman2000a}
Ofman, L., Romoli, M., Poletto, G., 2000a, ApJ, 529, 529

\bibitem[\protect\citeauthoryear{Ofman, Nakariakov \& DeForest}{1999}]{Ofman1999}
Ofman, L., Nakariakov, V.~M., DeForest, C.~E., 1999, ApJ, 514, 441.

\bibitem[\protect\citeauthoryear{Orrall \& Zirker}{1961}]{Orrall1961}
Orrall, F.~Q.,Zirker, J.~B., 1961, ApJ, 134, 72.

\bibitem[\protect\citeauthoryear{Papagiannis \& Baker}{1982}]{Papagiannis1982}
Papagiannis M.D. \& Baker K.B., 1982, SP, 79, 365.

\bibitem[\protect\citeauthoryear{Parnell}{2002a}]{Parnell2002a}
Parnell C.E., 2002a, COSPAR-CS 13, 47.

\bibitem[\protect\citeauthoryear{Parnell}{2002b}]{Parnell2002b}
Parnell, C.~E., 2002b, ESA SP-505, 231.

\bibitem[\protect\citeauthoryear{Pekeris}{1948}]{Pekeris1948}
Pekeris, C.~L., 1948, Geol.Soc.Amer.Mem.,  27, 117.

\bibitem[\protect\citeauthoryear{Pikel’ner}{1971}]{Pikelner1971}
Pikelner, S.~B., 1971, SP, 17, 44.

\bibitem[\protect\citeauthoryear{Roberts \& Joarder}{1994}]{Roberts1994}
Roberts, B., Joarder, P.~S., 1994, Proc-1994-Belvedere, 173.

\bibitem[\protect\citeauthoryear{Rosenberg}{1970}]{Rosenberg1970}
Rosenberg, H., 1970, A\&A, 9, 159.

\bibitem[\protect\citeauthoryear{Ryutova \& Tarbell}{2000}]{Ryutova2000}
Ryutova, M., Tarbell, T.~D., 2000, ApJL, 541, L29.

\bibitem[\protect\citeauthoryear{Ryutova, Habbal \& Woo}{2001}]{Ryutova2001}
Ryutova, M., Habbal, S., Woo, R.,  2001, SP, 200, 213.

\bibitem[\protect\citeauthoryear{Shimizu}{2002a}]{Shimizu2002a}
Shimizu, T., 2002a, COSPAR-CS, 13, 29.

\bibitem[\protect\citeauthoryear{Shimizu}{2002b}]{Shimizu2002b}
Shimizu, T., 2002b, ApJ, 574, 1074.

\bibitem[\protect\citeauthoryear{Shimizu, Tsuneta \& Acton}{1992}]{Shimizu1992}
Shimizu, T., Tsuneta, S.,  Acton, L.~W.,  1992, PASJ, 44, L147.

\bibitem[\protect\citeauthoryear{Shimizu, Tsuneta \& Acton}{1994}]{Shimizu1994}
Shimizu, T., Tsuneta, S., Acton, L.~W.,  1994, ApJ, 422, 906.

\bibitem[\protect\citeauthoryear{Shimizu}{1995}]{Shimizu1995}
Shimizu, T., 1995, PASJ, 47, 251.

\bibitem[\protect\citeauthoryear{Shimizu}{1997}]{Shimizu1997a}
Shimizu, T., 1997, PhD Thesis

\bibitem[\protect\citeauthoryear{Shimizu \& Tsuneta}{1997}]{Shimizu1997b}
Shimizu, T., Tsuneta, S., 1997, ApJ, 486, 1045.

\bibitem[\protect\citeauthoryear{Spadaro}{1999}]{Spadaro1999}
 Spadaro, D., 1999, ESA SP, 448, 157.

\bibitem[\protect\citeauthoryear{Spadaro, Noci \& Zappala}{1990a}]{Spadaro1990a}
Spadaro, D., Noci, G., Zappala, R.~A.,  1990a, ApJ, 355, 342.

\bibitem[\protect\citeauthoryear{Spadaro, Noci \& Zappala}{1990b}]{Spadaro1990b}
Spadaro, D., Noci, G., Zappala, R.~A., 1990b, ApJ, 362, 370.

\bibitem[\protect\citeauthoryear{Spicer, Benz \& Huba}{1981}]{Spicer1981}
Spicer, D.~S., Benz, A.~O., Huba, J.~D., 1981, A\&A 105, 221

\bibitem[\protect\citeauthoryear{Tanaka \& Papadopoulos}{1983}]{Tanaka1983}
Tanaka, K., Papadopoulos, K., 1983, Phys. of Fluids, 26, 1697.

\bibitem[\protect\citeauthoryear{Tarbell, Ryutova \& Covington}{1999}]{Tarbell1999}
Tarbell, T.~D., Ryutova, M., Covington, J., 1999, ApJ, 514, L47.

\bibitem[\protect\citeauthoryear{Tarbell, Ryutova \& Shine}{2000}]{Tarbell2000}
Tarbell, T.~D., Ryutova, M., Shine, R., 2000, SP, 193, 195.

\bibitem[\protect\citeauthoryear{Thejappa}{1987}]{Thejappa1987}
Thejappa, G., 1987, SP, 111, 45.

\bibitem[\protect\citeauthoryear{Timothy, Krieger \& Vaiana}{1975}]{Timothy1975}
Timothy, A.~F., Krieger, A.~S., Vaiana, G.~S., 1975, SP, 42, 135.

\bibitem[\protect\citeauthoryear{Tripathi, Grebogi \& Liu}{1977}]{Tripathi1977}
Tripathi, V.~K., Grebogi,C., Liu, C.~S., 1977, Phys. Fluids, 20, 1525

\bibitem[\protect\citeauthoryear{Vlahos}{1982}]{Vlahos1982}
Vlahos, L., Gergely, T.~E., Papadopoulos, K., 1982, ApJ, 258, 812.

\bibitem[\protect\citeauthoryear{Vourlidas, Bastian \& Aschwanden}{1997}]{Vourlidas1997}
Vourlidas A., Bastian T.S.\& Aschwanden M.J. 1997, ApJ 489, 403.

\bibitem[\protect\citeauthoryear{Wang, Schmahl \& Kundu}{1987}]{Wang1987}
Wang, Z., Schmahl, E.~J., Kundu, M.~R., 1987, SP, 111, 419.

\bibitem[\protect\citeauthoryear{Watari, Kozuka \& Ohyama}{1995}]{Watari1995}
Watari, S., Kozuka, Y., Ohyama, M., 1995, J.Geomag.Geoelec., 47(11), 1063.

\bibitem[\protect\citeauthoryear{White, Kundu \&  Shimizu}{1995}]{White1995}
White, S.~M., Kundu, M.~R., Shimizu, T., 1995, ApJ, 450, 435.

\bibitem[\protect\citeauthoryear{Zirker}{1993}]{Zirker1993}
Zirker, J.~B., 1993, SP, 148, 43-60.

\end{thebibliography}
\end{document}